\documentstyle[proceedings,numreferences,epsfig]{maxent95}

\def\Re{{\rm I\!R}}

\newcommand{\La}{\mbox{$\cal L\ $}}

\newcommand{\h}{\mbox{${ h}\ $} }

\newcommand{\p}{\partial}

\newcommand{\qed}{\mbox{{\bf Q.E.D.}}}
\newcommand{\proof}{\mbox{{\bf Proof} }}
\newcommand{\pa}{\mbox{${\mathrm pa}$}}
\newcommand{\an}{\mbox{${\mathrm an}$}}
\newcommand{\ap}{\mbox{${\mathrm ap}$}}
\newcommand{\wj}{\mbox{$\omega_{j}$}}
\newcommand{\woj}{\mbox{$\omega_{j}^{o}$}}
\newcommand{\xpa}[1]{\mbox{$x_{{\mathrm pa}(#1)}$}}
\newcommand{\xtpa}[1]{\mbox{$x^{(t)}_{{\mathrm pa}(#1)}$}}
\newcommand{\xap}[1]{\mbox{$x_{{\mathrm ap}(#1)}$}}

\newtheorem{theorem}{Theorem}

\newtheorem{lemma}{Lemma}

\begin{opening}
\title{Entropic Priors for Discrete Probabilistic Networks and for Mixtures of Gaussians Models}
\author{C. C. RODRIGUEZ}
\institute{Department of Mathematics and Statistics\\
University at Albany, SUNY\\
Albany NY 12222, USA\\
{\tt carlos@math.albany.edu\\
     http://omega.albany.edu:8008/}}
\end{opening}

\runningtitle{Entropic Priors for BBNs and Mixtures}

\begin{document}

\begin{abstract}
  The ongoing unprecedented exponential explosion of available
  computing power, has radically transformed the methods of 
  statistical inference. What used to be a small minority of
  statisticians advocating for the use of priors and a strict
  adherence to bayes theorem, it is now becoming the norm across
  disciplines.  The evolutionary direction is now clear. The trend is
  towards more realistic, flexible and complex likelihoods
  characterized by an ever increasing number of parameters. This makes
  the old question of: {\it What should the prior be?} to acquire a new
  central importance in the modern bayesian theory of inference.
  Entropic priors provide one answer to the problem of prior
  selection.  The general definition of an entropic prior has existed
  since 1988 \cite{kn:rdrgz89}, but it was not until 1998
  \cite{rodriguez98b} that it was found that they provide a new notion
  of complete ignorance. This paper re-introduces the family of
  entropic priors as minimizers of mutual information between the data
  and the parameters, as in \cite{rodriguez98b}, but with a small change
  and a correction. The general formalism is then applied to two 
large classes of models: Discrete probabilistic networks and univariate
finite mixtures of gaussians. It is also shown how to perform inference
by efficiently sampling the corresponding posterior distributions.

\end{abstract}

%\tableofcontents 
\section{Introduction}
Entropic Priors \cite{kn:rdrgz89,kn:skllng89,kn:rdrgz90,rodriguez91} minimize
a type of mutual information between the data and the parameters
\cite{rodriguez98b}. Hence, Entropic Priors are the prior models
that are most ignorant about the data. As Jaynes used to say: {\it
  they are maximally noncommittal with respect to missing information}.
Entropic Priors (as opposed to other prior assignments of probability)
come with a guarantee: They include only the information in the
likelihood, the initial guess, the hyper-parameter and the possible
side conditions that are explicitly imposed, and nothing else.
Entropic Priors provide a general recipe for prior probabilities that
allow the enjoyment of the bayesian omelet even in high dimensional
parameter spaces.

This paper presents the explicit computation of Entropic Priors for
two classes of models: General Discrete Probabilistic Networks (a.k.a.
Belief Nets, Bayesian Nets, BBNs) and for Mixtures of Gaussians
Models.  These models constitute the core of the probabilistic
treatment of uncertainty in AI.

The paper is divided into 5 parts. Section 2, repeats the derivation
in \cite{rodriguez98b} (but with a small change and a correction)
that Entropic Priors minimize mutual information between the data and
the parameters. Section 3, presents the computation for discrete BBNs.
Section 4 shows an application for classification. Section 5
computes the priors for the Mixture of Gaussians case. Finally some
general remarks and conclusions are included in Section 6.

\section{Entropic Priors are Most Ignorant Priors}
Given a regular parametric hypothesis space, i.e. a Riemannian
manifold of dominated probability distributions with volume element
$g^{1/2}(\theta)d\theta$. Where $g(\theta)$ is the determinant of the
Fisher information at $\theta$. We denote by $f(x|\theta)$ the density
(with respect to either Lebesgue or counting measure) of the
distribution indexed by $\theta$ and by $\pi(\theta)$ a prior density
on the parameters $\theta$. The entropic prior is the $\pi$ that makes
the joint distribution
\begin{equation}
  \label{eq:1}
  f(x_{1},\ldots,x_{\alpha},\theta) = \pi(\theta) \prod_{j=1}^{\alpha}
      f(x_{j}|\theta)
\end{equation}
hardest to discriminate (in the sense of minimizing the Kullback number)
from the independent model,
\begin{equation}
  \label{eq:2}
  h(x_{1},\ldots,x_{\alpha}) c g^{1/2}(\theta) \propto g^{1/2}(\theta)
      \left\{\prod_{j=1}^{\alpha} h(x_{j})\right\}
\end{equation}
for a given fix density $h(x)$ on the data space. Where $c$ is a
normalization constant independent of $\theta$ and the $x_{j}$s.
Notice that $c>0$ when the parameter space has finite volume.
However, the solution to the optimization problem (\ref{eq:5}) (and
hence, the entropic prior) does not depend on $c$ and still makes
sense for models with infinite volume. Notice further that the setting
is coherent in the sense that the rhs of (\ref{eq:2}) is in fact
proportional to the density of the model that assigns probabilities to
the $x_{\j}$s according to $h$ and independently of the $\theta$ which,
according with (\ref{eq:2}), is uniform over the surface area of the
model. This is true since Fisher information in the hypothesis space
of $\alpha$ independent observations is $\alpha$ times the Fisher
information in the hypothesis space of one observation and thus the
volume element in the space of $\alpha$ observations is $\alpha^{k/2}
g^{1/2}(\theta)$.  i.e., the two volume elements are proportional and
we assume the proportionality constant is included in $c$.

To simplify the notation
let $x^{\alpha} = (x_{1},\ldots,x_{\alpha})$ and write,
\begin{equation}
  \label{eq:3}
  I(\theta:h) = \int f(x|\theta) \log \frac{f(x|\theta)}{h(x)} dx
\end{equation}
and 
\begin{equation}
  \label{eq:4}
  I(f\pi:h g^{1/2}) = \int f(x^{\alpha}|\theta)\pi(\theta) 
    \log \frac{f(x^{\alpha}|\theta)\pi(\theta)}
    {h(x^{\alpha})c g^{1/2}(\theta)} dx^{\alpha} d\theta
\end{equation}
We have,
\begin{theorem}
  \label{th:1}
  \begin{equation}
    \label{eq:5}
  \pi^{*} = \mathop{\mbox{\rm argmin}}_{\pi} I(f\pi:h g^{1/2})    
  \end{equation}
where the minimum is taken over all the proper priors on the parameter
space, is given by the entropic prior:
\begin{equation}
  \label{eq:6}
  \pi^{*}(\theta | \alpha,h) \propto e^{-\alpha I(\theta:h)} g^{1/2}(\theta)
\end{equation}
\end{theorem}
\proof Using Fubbini's theorem, (\ref{eq:1}),(\ref{eq:2}) and the fact
that $\pi$ integrates to one, we can write
\begin{equation}
  \label{eq:7}
  I(f\pi:h g^{1/2}) = \alpha \int \pi(\theta) I(\theta:h) d\theta
    + \int \pi(\theta) \log \frac{\pi(\theta)}{g^{1/2}(\theta)} d\theta 
    - \log c.
\end{equation}
Therefore using a Lagrange multiplier to enforce the normalization
constraint ($\int \pi = 1$) we can find $\pi^{*}$ by solving:
\begin{equation}
  \label{eq:8}
\mathop{\mbox{\rm argmin}}_{\pi}  
  \int\left\{\alpha \pi(\theta) I(\theta:h) 
    + \pi(\theta) \log \frac{\pi(\theta)}{g^{1/2}(\theta)} 
    + \lambda \pi(\theta) \right\} d\theta 
\end{equation}
Let $\La(\pi,\lambda)$ denote the expression inside the curly
brackets in (\ref{eq:8}). The Euler-Lagrange equation for the optimal
$\pi^{*}$ is $\frac{\p\La}{\p\pi}=0$ given by,
\begin{equation}
  \label{eq:9}
  \alpha I + \log\pi^{*} - \log g^{1/2} + \lambda + 1 = 0.
\end{equation}
From where we obtain the expression for the entropic prior given
by (\ref{eq:6}). 
\\
\qed

\subsection{But What Does It Mean?}
First of all it needs to be clear that the above analysis is logically
a priori. By this I mean that the actual numerical values of the
observed data are not used, nor is the actual sample size number $n$
of observed $i.i.d.$ data vectors used. The parameters $\alpha$ and
$h$ of the entropic prior are the carriers of prior information.
Notice also that, since the derivation was done on a {\it virtual} and
not actual space of $\alpha$ observations, it makes sense to allow
$\alpha$ to take non integer values as long as $\alpha > 0$. In fact
an irrational $\alpha'$ is immediately obtained if we decide to change
(in the final formula for the entropic prior) the entropy scale to
{\it bits} by changing the original base of the logarithm in
$I(\theta:h)$ from $e$ to $2$ so that $\alpha' = \alpha \log 2$. It is
however incorrect to claim that by starting the derivation with
another base for the logarithm one will end up with a non integer
$\alpha'$ as it was wrongly claimed in \cite{rodriguez98b}.  In fact
the objective functions are proportional and they obviously produce
the same $\pi^{*}$. To see the source of the mistake one just needs to
notice that when the base of the $\log$ in (\ref{eq:9}) is $2$ say,
one has to exponentiate $2$, and not $e$, in order to solve for
$\pi^{*}$.  This was first pointed out to me by Ariel Caticha, who
then tried to build a justification for an entropic prior with fix
$\alpha=1$ in \cite{caticha2001}.

\subsubsection{Imaginary $\alpha$}
Allowing $\alpha$ to be not just a real number but a Clifford number,
in particular to be a pseudo scalar, opens up a garden of unexplored
possibilities. This may not be as insane as it first appears to be, if
one thinks of the resulting prior as the density of a Clifford valued
probability measure (see \cite{rodriguez98a}). Moreover, if $I$
(entropy) could be justified as $S$ (action) then the resulting {\it
  prior} $e^{i S/\hbar}$ (relative to local ignorance) would take a
familiar form.  Going with the flow of this (for now) applied
numerology this would point to current physical theory to be based on
the order of $10^{66}$ equivalent a priori observations! (i.e.
expressing $\hbar$ in geometrized units).

\subsection{Recipes for Choosing $\alpha$ and $\h$}
The values of the hyperparameters $\alpha$ and $h$ of the entropic
prior need to be fixed in order to obtain numerical assignments of
probabilities.  To fix $h$ we need to specify a function (i.e. an a
priori density $h(x)$ for the data) which involves, in principle, the
specification of an infinite number of parameters. Nevertheless, the
importance of the a priori biases introduced by $h$ are modulated by
the value of the real positive parameter $\alpha$. Take $\alpha$
sufficiently close to $0$ and the prior will be blind to the specific
form of $h$ and controlled by the volume element $g^{1/2}d\theta$
(i.e.  uniform over the model surface, see \cite{rodriguez93}).  There
is a close similarity with the problem of choosing a kernel and a
bandwidth in density estimation. As it is the case in density
estimation, the specific form of the kernel is not as critical as the
choice of the smoothness parameter. A natural choice for $h$ is to use
$h(x)=f(x|\theta_{0})$ where $\theta_{0}$ is the best current guess
for the value of $\theta$. If we assume the value of $\theta_{0}$ to
be unknown then we can consider the entropic prior model, which is now
indexed by the $1+k$ parameters $(\alpha,\theta_{0})$, to be another
regular hypothesis space that needs a prior on its parameters.  The
entropic prior on the entropic prior, on the entropic prior,$\ldots$,
etc is, in principle, computable. The possibility of a chain of
entropic priors for $\alpha$ was first given to first level in
\cite{kn:rdrgz89} and for all levels in \cite{kn:rdrgz90}. Another
general alternative is to use the empirical bayes approach (see
\cite{rodriguez91}). Finally, just fixing $\alpha$ to an arbitrary
small value ($\approx 1$) and using $\hat{\theta_{0}}$ the mle
(maximum likelihood estimator) or MAP (Maximum A posteriori
Probability), with an easy to handle conjugate prior, for $\theta$ has
been shown to perform well in simulation experiments.

\section{The Entropic Prior of a Discrete Probabilistic Network}
An understanding of Cox's \cite{cox46} argument should be sufficient to
impose the rules of probability to the treatment of uncertainty in AI.
But it has taken, however, a long heated debate (see \cite{cheeseman85}
and \cite{cheeseman88}), the invention of new efficient methods of
computation (e.g. the junction tree algorithm, see \cite{pnets99}) and
the publication of Pearl's text \cite{pearl88}, to arrive at today's
dominant view of a complete probabilistic approach.

\subsection{DAGs}
The current recipe for the thinking machine consists of a fully
bayesian probabilistic treatment of a long vector of facts (the data).
The main approach for encoding prior information about an specific
domain of application, is not the prior, but the likelihood. An a priori
network of conditional independence assumptions is typically provided by
means of a Directed Acyclic Graph (DAG) that is supposed to encode 
an expert's knowledge of causal relations among observable facts.

\begin{figure}
\begin{center}
\epsfysize=3.0in
\epsfbox{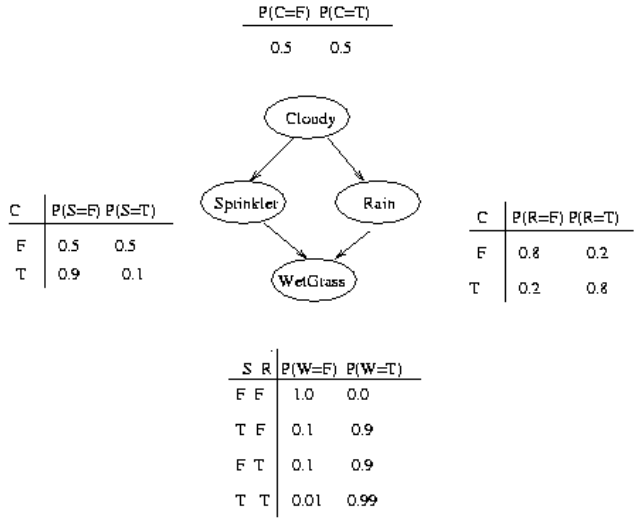}
\end{center}
\caption{DAG for the Sprinkler Problem}
\label{sprinkler}
\end{figure}

The canonical textbook example is displayed in fig~\ref{sprinkler}.
The arrows indicate causality. Thus, the presence of the arrow from
{\it Cloudy} to {\it Rain} represents the fact that the sky being
cloudy is a possible cause for rain. More important is the absence
of arrows which indicate independence. Thus, the picture shows
that conditionally on the values of {\it Sprinkler} and {\it Rain},
{\it Cloudy} is independent of {\it WetGrass}. The entries of the tables
of conditional probabilities constitute the parameters of the DAG.
In the case of fig~\ref{sprinkler} there are $9$ independent parameters.
We can think of a DAG as a convenient way to specify a high dimensional
submanifold of the space of all joint distributions of the variables under
consideration. For example, the pictured DAG (with unspecified tables)
represents a $9$ dimensional submanifold of the $15$ dimensional simplex
of all the assignments of probability on the $2^{4}=16$ possible observations
of the binary variables $(C,S,R,W)$. The DAG in fig~\ref{sprinkler} 
specifies the joint distribution of all the variables $(C,S,R,W)$ in terms
of the parameters $\theta$ (i.e. table entries) as,
\begin{equation}
  \label{eq:10}
  P(C,R,S,W) = P(C)P(R|C)P(S|R)P(W|R,S).
\end{equation}
Each of the factors on the right of equation (\ref{eq:10}) can be read off
the tables provided in fig~\ref{sprinkler}. For example, 
\begin{equation}
  \label{eq:11}
  P(C=T,R=T,S=F,W=F) = (0.5)(0.8)(0.5)(0.1) = 0.02
\end{equation}
In order to provide general formulas for DAGs we number the vector of
variables by $x=(x_{1},x_{2},x_{3},x_{4})=(C,R,S,W)$ and parameterized the
joint distribution with a vector $\theta$ of parameters as in,
\begin{equation}
  \label{eq:12}
  \theta_{4w}(r,s) = P(W=w|R=r,S=s)
\end{equation}
Thus, labeling $F=1$ and $T=2$, (\ref{eq:11}) becomes,
\begin{equation}
  \label{eq:13}
  P(2,2,1,1|\theta) = \theta_{12}\theta_{22}(2)\theta_{31}(2)\theta_{41}(2,1)
\end{equation}

\begin{figure}
\begin{center}
\epsfysize=2.0in
\epsfbox{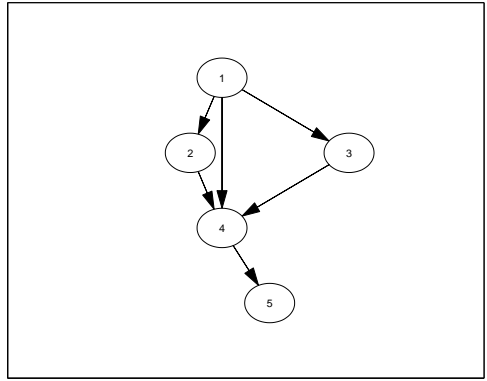}
\end{center}
\caption{Example of a DAG}
\label{adag}
\end{figure}

\subsection{Who is Who on a DAG: General Notation}
This section provides some definitions and notations that are needed
for writing the entropic prior on a general DAG. All the examples
refer to fig~\ref{adag}.

\begin{description}

\item[Directed Graph:] An ordered pair $(V,E)$ where $V$ is a set of vertices 
(e.g. $V = \{1,2,3,4,5\}$) and $E \subset V \times V$ is a set of directed 
edges. e.g., 
\[E = \{(1,2), (1,3), (1,4), (2,4), (3,4), (4,5)\}\]

\item[DAG:] A Directed Acyclic Graph is a directed graph without cycles.
(e.g. fig~\ref{adag}).

\item[Parents:] $\pa(k)$ denotes the set of parents for the vertices $k\in V$.
(e.g. $\pa(1)=\phi, \pa(5)=\{4\}, \pa(4)=\{1,2,3\}$).

\item[Ancestors:] $\an(k)$ denotes the set of ancestors of $k\in V$.
(e.g. $\an(2)=\{1\}, \an(5)=\{1,2,3,4\}, \an(1)=\phi$). Clearly,
\begin{equation}
\label{eq:14}
 \an(k) = \pa(k) \bigcup_{j\in \pa(k)} \an(j) 
\end{equation}

\item[Ancestors that are not Parents:] Denoted by $\ap(k)$
  \begin{equation}
    \label{eq:14.5}
    \ap(k) = \an(k)\setminus \pa(k)
  \end{equation}
(e.g. $\ap(5) = \{1,2,3\}, \ap(4) = \phi, \ap(2)=\phi$).

\item[Notation:] 
  \begin{equation}
    \label{eq:15}
    \xpa{k} \equiv \left\{x_{j} : j\in \pa(k) \right\}
  \end{equation}
e.g. 
\[ \xpa{1}=\phi, \ \xpa{4} = \{ x_{1},x_{2},x_{3} \} \]

\item[Notation:] $\displaystyle\sum_{\xpa{k}}$ denotes the multiple
  sum over all the possible values of the variables that are parents
  of vertice $k \in V$. e.g.

\[ \sum_{\xpa{4}} \equiv \sum_{x_{1}}\sum_{x_{2}}\sum_{x_{3}} \]

\end{description}

The notation introduced with equation (\ref{eq:13}) generalizes
naturally for any number of discrete variables.  Given a DAG with set
of vertices $V$ we let $x=\{x_{k}: k\in V\}$. Hence, the joint
distribution of the variables of a given DAG is given by,
\begin{eqnarray}
  \label{eq:16}
  p(x|\theta) &=& \prod_{k\in V} p(x_{k}|\xpa{k},\theta) \nonumber \\
              &=& \prod_{k\in V} \theta_{k x_{k}}(\xpa{k})
\end{eqnarray}
We are now ready to compute.

\subsection{Entropy of a DAG}
Given a DAG, the Kullback number between two sets of parameters $\theta$
and $\mu$ is,
\begin{equation}
  \label{eq:17}
  I(\theta : \mu) = E_{\theta}\left[\log\frac{p(x|\theta)}{p(x|\mu)}\right]
\end{equation}
Using (\ref{eq:16}) and interchanging expectation with summation we obtain,
\begin{equation}
  \label{eq:18}
  I(\theta:\mu) = \sum_{k\in V} E_{\theta}\left[
    \log\frac{\theta_{k x_{k}}(\xpa{k})}{\mu_{k x_{k}}(\xpa{k})} \right]
\end{equation}
Now for each $k\in V$ compute the unconditional expectation in (\ref{eq:18})
by first conditioning on the values of $\xpa{k}$ to obtain,
\begin{eqnarray}
  \label{eq:19}
  E_{\theta}\left[\left.
    \log\frac{\theta_{k x_{k}}(\xpa{k})}{\mu_{k x_{k}}(\xpa{k})}
          \right|\xpa{k}\right] &=&
        \sum_{j=1}^{r_{k}} \theta_{kj}(\xpa{k})
    \log\frac{\theta_{kj}(\xpa{k})}{\mu_{kj}(\xpa{k})} \nonumber \\
    &=& I(\theta_{k}(\xpa{k}):\mu_{k}(\xpa{k}))
\end{eqnarray}
where the last equality is a definition and it was assumed that
$x_{k}$ can take $r_{k}$ discrete values. Taking expectations over
the $\xpa{k}$ and replacing in (\ref{eq:18}) we obtain,
\begin{equation}
  \label{eq:20}
  I(\theta:\mu) = \sum_{k\in V} \sum_{\xpa{k}} p(\xpa{k}|\theta)\ 
               I\left(\theta_{k}(\xpa{k}):\mu_{k}(\xpa{k})\right).
\end{equation}
Finally, using the fact that,
\begin{eqnarray}
  \label{eq:21}
  p(\xpa{k}|\theta) &=& \sum_{\xap{k}}p(\xap{k},\xpa{k}|\theta) \nonumber \\
  &=& \sum_{\xap{k}} \prod_{j\in \an(k)}p(x_{j}|\xpa{j},\theta) \nonumber \\
  &=& \sum_{\xap{k}} \prod_{j\in \an(k)} \theta_{j x_{j}}(\xpa{j})
\end{eqnarray}
we obtain the expression for the entropy,
\begin{equation}
  \label{eq:22}
  I(\theta:\mu) = \sum_{k\in V} \sum_{\xpa{k}}
     \left\{
       \sum_{\xap{k}} \prod_{j\in \an(k)} \theta_{j x_{j}}(\xpa{j})
     \right\}\ I(\theta_{k}(\xpa{k}):\mu_{k}(\xpa{k})).
\end{equation}
Thus, formula (\ref{eq:20}) shows that the total entropy for a DAG
is obtained by adding the entropies for each node. The entropy of
a node is computed as an average of all the possible entropies obtained
for the different values of the parents of that node. 
In practice formula (\ref{eq:22}) may be too expensive to compute and
it may be necessary to use a Monte Carlo estimate.

\subsection{Volume Element of a DAG}
To compute the Fisher metric, write $\theta$ as a long vector and
use the fact (see \cite{rodriguez91}) that,
\begin{equation}
  \label{eq:23}
  I(\theta:\theta+\epsilon v) = \frac{\epsilon^{2}}{2}
     \sum_{i,j} g_{ij}(\theta)v^{i} v^{j} + o(\epsilon^{2})
\end{equation}
It then follows immediately from (\ref{eq:22}) that the Fisher matrix
is block diagonal. Each block corresponds to the $(r_{k}-1)\times
(r_{k}-1)$ (Fisher matrix $G_{k}(\theta_{k}(\xpa{k}))$ associated to
the $k$th node, multiplied by the scalar $p(\xpa{k}|\theta)$. The
determinant, $g(\theta)$, of the Fisher matrix is then given by the
product of the determinants of each of the blocks. We have
\begin{equation}
  \label{eq:24}
  g(\theta) = \prod_{k\in V}\prod_{\xpa{k}} 
     \left\{
       \sum_{\xap{k}} \prod_{j\in \an(k)} \theta_{j x_{j}}(\xpa{j})
     \right\}^{r_{k}-1} \ \det G_{k}\left(\theta_{k}(\xpa{k})\right)
\end{equation}
Finally using the fact that $G_{k}$ is the Fisher matrix of a 
multinomial with parameters $\theta_{k1}(\xpa{k}), \ldots, 
 \theta_{k r_{k}}(\xpa{k})$ we have,
 \begin{equation}
   \label{eq:25}
   \det G_{k}\left(\theta_{k}(\xpa{k})\right) = 
     \frac{1}{\displaystyle\prod_{j=1}^{r_{k}}\theta_{kj}(\xpa{k})}
 \end{equation}
replacing (\ref{eq:25}) in (\ref{eq:24}) and taking square root we obtain
the expression for the volume element,
\begin{equation}
  \label{eq:26}
  g^{1/2}(\theta)\ d\theta  = \prod_{k\in V}\prod_{\xpa{k}} 
     \frac{\displaystyle \left\{
       \sum_{\xap{k}} \prod_{j\in \an(k)} \theta_{j x_{j}}(\xpa{j})
     \right\}^{(r_{k}-1)/2}}
   {\displaystyle \prod_{j=1}^{r_{k}}\theta^{1/2}_{kj}(\xpa{k})}\ d\theta
\end{equation}

\subsection{The Entropic Prior for a DAG}
To obtain (\ref{eq:6}) we use (\ref{eq:22}), (\ref{eq:21}) and
(\ref{eq:26}) to get,
\begin{eqnarray}
  \label{eq:27}
  \pi(\theta|\alpha,\mu) \propto &\ & \prod_{k\in V}\prod_{\xpa{k}}
   \left\{\frac{p^{(r_{k}-1)}(\xpa{k}|\theta)}
   {\displaystyle \prod_{j=1}^{r_{k}}\theta_{kj}(\xpa{k})}\right\}^{1/2}
 \nonumber \\
  & & \exp\left\{ -\alpha p(\xpa{k}|\theta)
   I(\theta_{k}(\xpa{k}):\mu_{k}(\xpa{k})) \right\}
\end{eqnarray}

\subsection{Posterior}
Let us assume that there is available a set of $N$ independent observations
\begin{equation}
  \label{eq:030}
  D = \{ x^{(1)}, x^{(2)}, \ldots, x^{(N)} \}
\end{equation}
where each $x^{(t)}=(x^{(t)}_{1},\ldots,x^{(t)}_{n})$ is an $|V|=n$ dimensional
vector containing the observed values of the nodes of a general DAG $(V,E)$.
As usual the posterior is given by Bayes theorem as,
\begin{equation}
  \label{eq:031}
  \pi(\theta|D,\alpha,\mu) \propto f(D|\theta)\  \pi(\theta|\alpha,\mu)
\end{equation}
where the likelihood is given by,
\begin{eqnarray}
  \label{eq:032}
  f(D|\theta) &=& \prod_{t=1}^{N} f(x^{(t)}|\theta) \nonumber \\
  &=& \prod_{t=1}^{N} \prod_{k=1}^{n} 
    f(x^{(t)}_{k}|\xtpa{k},\theta) \nonumber \\
  &=& \prod_{t=1}^{N} \prod_{k=1}^{n}
      \theta_{k x^{(t)}_{k}}(\xtpa{k})
\end{eqnarray}
Let us partition the set of vertices into two groups, those with parents 
and those without (orphans). For the orphan nodes, i.e. for $k\in V$ 
such that $\pa(k)=\phi$ and  for $i=1,2,\ldots, r_{k}$ define
\begin{equation}
  \label{eq:033}
  n_{ki}(\phi) = \left| \{t : x^{(t)}_{1} = i \} \right|
\end{equation}
and for $k\in V$ with $\pa(k)\ne \phi$ and $i=1,2, \ldots, r_{k}$ 
\begin{equation}
  \label{eq:034}
  n_{ki}(\xpa{k}) = \left| \{t : x^{(t)}_{k} = i \mbox{\ and\ } 
    \xtpa{k} = \xpa{k} \} \right|
\end{equation}
Replacing these counts into (\ref{eq:032}) we obtain,
\begin{equation}
  \label{eq:035}
  f(D|\theta) = \prod_{k=1}^{n}\prod_{\xpa{k}}\prod_{i=1}^{r_{k}}
  \left\{ \theta_{ki}(\xpa{k}) \right\}^{n_{ki}(\xpa{k})}
\end{equation}
To simplify the notation let us write simply by $p_{k}$ the expression
(\ref{eq:21}) which is always a probability that depends only on the
ancestors of the node $k$. Let us also just write $\theta_{ki},
n_{ki}, \mu_{ki}$ instead of $\theta_{ki}(\xpa{k}), \ldots$ and keep implicit
their dependence on given values of the parents. With this notation the
posterior becomes,
\begin{equation}
  \label{eq:0351}
  \pi(\theta|D,\alpha,\mu) \propto \prod_{k\in V}\prod_{\xpa{k}}
    p_{k}^{(r_{k}-1)/2} \prod_{i=1}^{r_{k}} \left\{
      \theta_{ki}^{n_{ki}-\frac{1}{2}}\ 
        \exp\left( -\alpha p_{k}\theta_{ki} \log 
          \frac{\theta_{ki}}{\mu_{ki}} \right)
        \right\}
\end{equation}
were we have used (\ref{eq:19}) to write the exponential in (\ref{eq:27})
as a product of $r_{k}$ factors.

\section{Example: Na\"{i}ve Bayes}
\begin{figure}
\begin{center}
\epsfysize=2.0in
\epsfbox{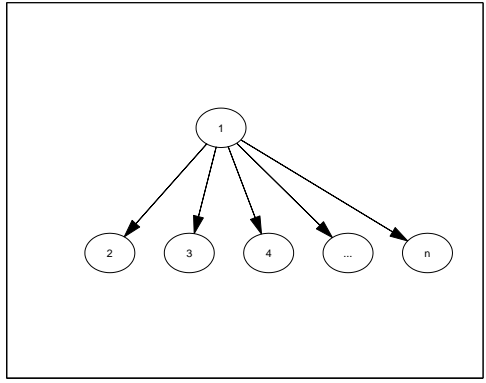}
\end{center}
\caption{DAG for Na\"{i}ve Bayes}
\label{naive}
\end{figure}
When the DAG has the form shown in fig~\ref{naive} the general formulas 
have simpler forms. This case is known as na\"{i}ve bayes
and it is often used as an approximation in discrimination problems.
For this case, $V= \{1,\ldots,n\}, \pa(1)=\phi$, and for $k\ne 1$ we
have $\pa(k)= \{1\}, \an(k)= \{1\}, \ap(k)= \phi$ and,
\begin{equation}
  \label{eq:027}
  p(\xpa{k}|\theta) = p(x_{1}|\theta_{1}) = \theta_{1 x_{1}}
\end{equation}
The expression for the entropy (\ref{eq:22}) becomes,
\begin{equation}
  \label{eq:28}
  I(\theta:\mu) = I(\theta_{1},\mu_{1}) + \sum_{k=2}^{n}\sum_{j=1}^{r_{1}}
    \theta_{1j}\ I(\theta_{k}(j):\mu_{k}(j))
\end{equation}
and the volume element (\ref{eq:26}) reduces to,
\begin{equation}
  \label{eq:29}
  g^{1/2}(\theta)\ d\theta = \frac{\displaystyle
    \left(\prod_{j=1}^{r_{1}} \theta_{1j}\right)^{(\sum_{k=2}^{n}r_{k}-1)/2}}
  {\displaystyle \left(
    \prod_{j=1}^{r_{1}}\prod_{k=2}^{n}\prod_{i=1}^{r_{k}}\theta_{ki}(j)
    \right)^{1/2}}\ d\theta
\end{equation}
The entropic prior is then easily computed by multiplying 
$\exp(-\alpha I(\theta:\mu))$ (obtained from (\ref{eq:28})) by (\ref{eq:29}).

\subsection{Posterior}
For na\"{i}ve bayes the likelihood is given by,
\begin{equation}
  \label{eq:32}
  f(D|\theta) = \prod_{i=1}^{N} \theta_{1 x^{(i)}_{1}} \prod_{k=2}^{n}
      \theta_{k x^{(i)}_{k}}(x^{(i)}_{1})
\end{equation}
Replacing the counts into (\ref{eq:32}) we obtain,
\begin{equation}
  \label{eq:35}
  f(D|\theta) = \left( \prod_{j=1}^{r_{1}}\theta_{1j}^{n_{1j}}\right)\ 
  \left( 
    \prod_{j=1}^{r_{1}}\prod_{k=2}^{n}\prod_{i=1}^{r_{k}}
    \left(\theta_{ki}(j)\right)^{n_{ki}(j)}
  \right)
\end{equation}
Letting,
\begin{equation}
  \label{eq:36}
  m = \frac{1}{2}\left(\sum_{k=2}^{n}r_{k}-n\right)
\end{equation}
we can write the posterior as,
\begin{eqnarray}
  \label{eq:37}
  & &\pi(\theta|D,\alpha,\mu) \propto \left\{ \prod_{j=1}^{r_{1}}
   \theta_{1j}^{m+n_{1j}-\alpha\theta_{1j}} \exp\left(
     -(\alpha\log \frac{1}{\mu_{1j}}) \theta_{1j}\right)\right\} \\
   & & \left\{
     \prod_{j=1}^{r_{1}}\prod_{k=2}^{n}\prod_{i=1}^{r_{k}}
       \left( \theta_{ki}(j) \right)^{n_{ki}(j)-\frac{1}{2}-
         \alpha\theta_{1j}\theta_{ki}(j)} \exp\left(
           -(\alpha\theta_{1j}\log \frac{1}{\mu_{ki}(j)}) \theta_{ki}(j)\right)
     \right\} \nonumber
\end{eqnarray}

\subsection{The Entropic Sampler}
A combination of Gibbs and Metropolis can be used for sampling the
posterior (\ref{eq:37}). The parameters are naturally grouped in
blocks $\theta_{k}$, where,
\begin{eqnarray}
  \label{eq:371}
  \theta_{k} &=& \theta_{k}(\xpa{k}) \nonumber \\
  &=& (\theta_{k1},\ldots, \theta_{kr_{k}}) \ \mbox{with\ } 
     \sum_{i=1}^{r_{k}} \theta_{ki} = 1
\end{eqnarray}
are distributed over the simplex of dimension $r_{k}-1$.
It can be readily seen from (\ref{eq:37}) that the marginal joint
distributions of the  $\theta_{k}$ blocks are all of the generic form,
\begin{equation}
  \label{eq:40}
  f(y_{1},y_{2},\ldots,y_{r-1}) \propto \prod_{j=1}^{r}\left\{
    y_{j}^{\alpha_{j}-1}\ e^{-\beta_{j} y_{j}}\right\}
\end{equation}
with $y_{j} \ge 0$ and $y_{r} = 1-\sum_{j=1}^{r-1}y_{j}$. The
parameters $\alpha_{j}$ and $\beta_{j}$ are different for the
parent node and for the children nodes. For the parent,
\begin{eqnarray}
  \label{eq:41}
  \alpha_{j} &=& 1+ m + n_{1j} - \alpha \theta_{1i} 
     \approx 1+m+n_{1j} \\
  \beta_{j} &=& \alpha \left(\log\frac{1}{\mu_{1j}}+
    \sum_{k=2}^{n} I(\theta_{k}(j):\mu_{k}(j))\right)
\end{eqnarray}
For the children blocks the parameters are,
\begin{eqnarray}
  \label{eq:42}
  \alpha_{j} &=& n_{ki}(j)+\frac{1}{2}-\alpha \theta_{1j}\theta_{ki}(j)
    \approx  n_{ki}(j)+\frac{1}{2} \\
  \beta_{j} &=& \alpha \theta_{1j} \log\frac{1}{\mu_{ki}(j)}
\end{eqnarray}
Excellent initial distributions for Metropolis are obtained
by using the following,
\begin{lemma}
  \label{lm:1}
  Let $y_{1},y_{2},\ldots, y_{r}$ be independent with $y_{j}$ following
a Gamma distribution with parameters $(\alpha_{j},\beta_{j})$. Let,
\begin{equation}
  \label{eq:43}
  z_{j} = \frac{y_{j}}{y_{1}+\cdots +y_{r}}\ \mbox{for\ } j=1,\ldots,r-1
\end{equation}
then the joint density of the $z_{j}$'s is given by,
\begin{equation}
  \label{eq:44}
  f(z_{1},\ldots,z_{r-1}) \propto \frac{\displaystyle \prod_{j=1}^{r}
    z_{j}^{\alpha_{j}-1}}{\displaystyle\left( \sum_{j=1}^{r}
      \beta_{j}z_{j}\right)^{\alpha_{1}+\alpha_{2}+\cdots +\alpha_{r}}}
\end{equation}
where $z_{r} \equiv 1-z_{1}-z_{2}-\cdots - z_{r-1}$.
\end{lemma}
\proof 
%Condition on $y_{r}$ and change the variables. After a long
%but straight forward simplification one arrives at (\ref{eq:44}).
Notice that (\ref{eq:44}) is a generalization of the classic result
for the Dirichlet distribution obtained when all the $\beta_{j}$'s are
equal, in which case the denominator becomes proportional to $1$. To
prove (\ref{eq:44}) just condition on $y_{r}=y$ so that the transformation
(\ref{eq:43}) from the $y_{j}$'s to the $z_{j}$'s for $j=1,2, \ldots,
r-1$ is one to one with inverse,
\begin{equation}
  \label{eq:44.5}
  y_{j} = \frac{y z_{j}}{z_{r}}\ \mbox{for\ } j=1,\dots,r-1
\end{equation}
To show (\ref{eq:44.5}) just notice that,
\begin{eqnarray}
  \label{eq:44.6}
  y\ z_{j} &=& \frac{y_{j}\ y}{y+\sum_{i=1}^{r-1}y_{i}} \\
  &=& y_{j} \left( 1 - \frac{\sum_{i=1}^{r-1}y_{i}}
    {y+\sum_{i=1}^{r-1}y_{i}}\right) \\
  &=& y_{j} \left( 1 - \sum_{i=1}^{r-1} z_{i} \right) \\
  &=& y_{j}\ z_{r}
\end{eqnarray}
where we have used (\ref{eq:43}) and the definition of $z_{r}$.
The probability density of observing $z_{1},\ldots, z_{r-1}$ is then,
\begin{equation}
  \label{eq:44.8}
  f(z_{1},\ldots,z_{r-1}) = \int_{0}^{\infty} f(z_{1},\ldots,z_{r-1}|y_{r}=y)
    g_{r}(y)\ dy
\end{equation}
where $g_{j}$ for $j=1,\ldots,r$ are the gamma densities of the $y_{j}$.
Using the definition of the $z_{j}$'s given in (\ref{eq:43}), the
assumed independence of the $y_{j}$'s, and the change of variables
theorem together with (\ref{eq:45}), we have,
\begin{equation}
  \label{eq:44.9}
  f(z_{1},\ldots,z_{r-1}|y_{r}=y) = \left(\prod_{j=1}^{r-1} 
    g_{j}\left(\frac{y\ z_{j}}{z_{r}}\right)\right)\ \frac{1}{z_{r}}
     \left(\frac{y}{z_{r}}\right)^{r-1}
\end{equation}
The expression outside the product is the determinant of the Jacobian
of the transformation (\ref{eq:44.5}). This can be seen by noticing
that the Jacobian matrix is,
\begin{equation}
  \label{eq:44.10}
J = \frac{y}{z_{r}^{2}}\ 
   \left [\begin {array}{cccc} z_{1}+z_{r}&z_{1}&\ldots&z_{1}
\\\noalign{\medskip} z_{2}&z_{2}+z_{r}&\ldots&z_{2}
\\\noalign{\medskip}  & &\ddots& 
\\\noalign{\medskip} z_{r-1}&z_{r-1}&\ldots&z_{r-1}+z_{r}
\end {array}
\right ]
\end{equation}
and compute its determinant by subtracting from each column the column
that follows, to obtain,
\begin{equation}
  \label{eq:44.11}
\det J = \left(\frac{y}{z_{r}^{2}}\right)^{r-1}\ 
   \left |\begin {array}{ccccc} z_{r}&0&\ldots&0&z_{1}
\\\noalign{\medskip} -z_{r}&z_{r}&\ldots&0&z_{2}
\\\noalign{\medskip}  & &\ddots& 
\\\noalign{\medskip} 0&0&\ldots&-z_{r}&z_{r-1}+z_{r}
\end {array}
\right |
\end{equation}
and expanding along the last column,
\begin{eqnarray}
  \label{eq:44.12}
  \det J &=& \left(\frac{y}{z_{r}^{2}}\right)^{r-1} \left(
   z_{1}z_{r}^{r-2} + z_{2}z_{r}^{r-2}+\ldots + z_{r-2}z_{r}^{r-2}
     + (z_{r-1}+z_{r})z_{r}^{r-2} \right) \nonumber \\
   &=& \left(\frac{y}{z_{r}^{2}}\right)^{r-1} z_{r}^{r-2} \nonumber \\
   &=& \frac{1}{z_{r}} \left(\frac{y}{z_{r}}\right)^{r-1}
\end{eqnarray}
This proves (\ref{eq:44.9}). Replacing (\ref{eq:44.9}) into
(\ref{eq:44.8}) and using the expressions for the gamma densities we
obtain,
\begin{equation}
  \label{eq:44.13}
  f(z_{1},\ldots,z_{r-1}) \propto 
    \left(\prod_{j=1}^{r}z_{j}^{\alpha_{j}-1}\right)\frac{1}{z_{r}}
      \int_{0}^{\infty} \left(\frac{y}{z_{r}}\right)^
        {\sum_{j=1}^{r}\alpha_{j}-1}
      \ \exp\left\{- \left(\frac{1}{z_{r}}\sum_{j=1}^{r}\beta_{j}z_{j}
          \right) y\right\}\ dy
\end{equation}
this is a simple gamma integral. Integrating out and simplifying the $z_{r}$'s
we obtain the desired result (\ref{eq:44}).\\
\qed

To generate approximate samples from (\ref{eq:40}) we use the Lemma but
with $\tilde{\beta_{j}}$ chosen so that,
\begin{equation}
 \label{eq:45}
 \frac{C}{\displaystyle \left(\sum_{j=1}^{r} \tilde{\beta_{j}} 
     z_{j}\right)^{\alpha_{1}+\cdots +\alpha_{r}}} \approx 
 \exp\left(-\sum_{j=1}^{r}\beta_{j}z_{j}\right)
\end{equation}
where the constant $C$ does not depend on the $z_{j}$.
To find the $\tilde{\beta_{j}}$ just write the left side of (\ref{eq:45})
in exponential form and use,
\begin{equation}
  \label{eq:46}
  \log(\tilde{\beta_{1}}z_{1}+\cdots + \tilde{\beta_{r}}z_{r}) = 
  \log(\tilde{\beta_{r}}) + \log\left(
 1+\frac{\tilde{\beta_{1}}-\tilde{\beta_{r}}}{\tilde{\beta_{r}}} z_{1} 
 + \cdots + \frac{\tilde{\beta}_{r-1}-\tilde{\beta_{r}}}{\tilde{\beta_{r}}} 
  z_{r-1} \right) 
\end{equation}
together with,
\begin{equation}
  \label{eq:47}
  \log(1+z) = z + o(z)
\end{equation}
we obtain, that in order for (\ref{eq:45}) to be true, we must have,
\begin{equation}
  \label{eq:48}
  \frac{\tilde{\beta}_{j}-\tilde{\beta}_{r}}{\tilde{\beta}_{r}}
   \sum_{i=1}^{r}\alpha_{i} = \beta_{j}-\beta_{r}
\end{equation}
we can then use,
\begin{eqnarray}
  \label{eq:49}
  \tilde{\beta}_{r} = \sum_{i=1}^{r}\alpha_{i} \\
  \tilde{\beta}_{i} = \beta_{i}-\beta_{r}+\tilde{\beta}_{r}
\end{eqnarray}
Metropolis corrections are needed to correct for the approximations
introduced in (\ref{eq:41}), (\ref{eq:42}) and (\ref{eq:47}).

\subsection{Test: Credit Card Classification Example}
We tested the performance of the MCMC sampler on a standard set of
$10000$ data records containing the $13$ variables in table~\ref{tb:1}.
\begin{table}[h]
  \centering
  \begin{tabular}{ll}
    Nodes & Sizes \\
    \hline
Card = C           &    2(4)\\
Gender = G         &    2\\
Country = Y        &    3\\
Age = A            &    9 \\
State = S          &    13\\
Education = E      &    5\\
Marital = M        &    2\\
Occupation = O     &    5 \\
Total children = T &    6\\
Income = I         &    8\\
House owner = H    &    2\\
Cars owned = R     &    5\\
Children home = N  &    6 \\
   \hline
  \end{tabular}
  \caption{Data Records in Example}
  \label{tb:1}
\end{table}
Most of the node names are self explanatory. Card, originally contained
the type of credit card owned by the individual with categories:
{\it no card, regular, gold, platinum}. These were later reduced to only
two categories: {\it \{no card, regular\} and \{gold, platinum\}}. The data
contains individuals from the three north american countries: 
{\it Mexico, US, Canada}. However, the majority of records are from the US.
The {\it Children home} variable contains information about the actual 
number of children living at home with the individual.

\subsubsection{The Bayes Classifier}
To test the performance of the entropic sampler we chose at random 100
individuals to be used as the observed data and 1000 to test the 
bayes classifier. The bayes classifier simply assigns the category
with highest posterior probability.

Let $D$ be the observed $N=100$ records and let $x_{2},\ldots, x_{n}$
(here $n=13$) be the values of all the nodes except the first (i.e.
{\it Card}) for an individual that we want to classify. The bayes
classifier allocates $x_{1}=1$ if,
\begin{equation}
  \label{eq:50}
  P\left(x_{1} = 1 | x_{2},\ldots, x_{n},D \right) >
      P\left(x_{1} = 2 | x_{2},\ldots, x_{n},D \right) 
\end{equation}
we compute both sides with,
\begin{eqnarray}
  \label{eq:51}
  P\left(x_{1} = j | x_{2},\ldots, x_{n},D \right)  &=&
    \int   P\left(x_{1} = j, \theta | x_{2},\ldots, x_{n},D \right)\ d\theta
    \nonumber \\
  &\propto& 
    \int   P\left(x_{1} = j, x_{2},\ldots, x_{n}, \theta |D \right)\ d\theta
    \nonumber \\
  &=& \int   P\left(x_{1} = j, x_{2},\ldots, x_{n}| \theta \right)\ 
   \pi(\theta|D)\ d\theta
\end{eqnarray}
where we have assumed that the values of the individual to be
classified are independent of the observed data $D$. We use the
MCMC sampler to estimate (\ref{eq:51}) for $j=1$ and $j=2$.
Thus, if the sampler produces $\theta^{(1)},\ldots , \theta^{(M)}$
samples from the posterior $\pi(\theta|D)$ we classify $x_{1}=1$
if,
\begin{equation}
  \label{eq:52}
  \sum_{t=1}^{M} p(1,x_{2},\ldots,x_{n}|\theta^{(t)}) >
      \sum_{t=1}^{M} p(2,x_{2},\ldots,x_{n}|\theta^{(t)})
\end{equation}
To avoid underflows it is better to use only ratios. A more stable rule
is then: assign $x_{1}=1$ if,
\begin{equation}
  \label{eq:53}
  \sum_{t=1}^{M} \left( 1 - \frac{p(2,x_{2},\ldots,x_{n}|\theta^{(t)})}
    {p(1,x_{2},\ldots,x_{n}|\theta^{(t)})} \right)
  \frac{p(1,x_{2},\ldots,x_{n}|\theta^{(t)})}
    {p(1,x_{2},\ldots,x_{n}|\theta^{(1)})} > 0
\end{equation}

\subsubsection{Preliminary Results}
Table~\ref{tb:2} shows the results of running the sampler with
different parameter values.
\begin{table}[h]
  \centering
  \begin{tabular}{lllllll}
    burn &  M  &  N  &  inter  &  Met  &  $\alpha$  &  \% succ. \\
    \hline
    100  &  100 & 100 & 50 & [30 15] & 10  & 82.7 \\
    200  &  100 & 100 & 100& [5 2]  & 0.1  & 81.2 \\
    1000 &  200 & 100 & 50 & [2 2] &  1.0  & 78.4 \\
    1000 &  200 & 100 & 100& [1 1] &  1.0  & 79.0 \\
    100  &  100 & 50  & 50 & [1 1] &  1.0  & 76.3 \\
    \hline
  \end{tabular}
  \caption{Summary of Simulations}
  \label{tb:2}
\end{table}
The burn column contains the number of complete sweeps performed and 
discarded before collecting samples. The other columns are: M the number
of thetas sampled, N the observed sample size, inter the number of
discarded sweeps between samples, Met is the number of metropolis
step corrections for the root node and for the children nodes, $\alpha$ is
the parameter of the entropic prior and finally, \% succ. is the
percentage of correct classifications on 1000 random tests.

Notice that the metropolis corrections seem to help but they slow
down the sampler. Notice also the drop in performance when the
sample size becomes 50. 

These results show the adequacy of the entropic sampler
for the classification task. However, the na\"{i}ve bayes DAG is
not competitive with DAGs containing more realistic structure for this 
problem. A simulated annealing search over the space of DAGs produces
structures showing over $84\%$ success rate in the more difficult
task of classification with 4 (not just 2) categories of credit card.

\section{Entropic Prior for Mixtures of Gaussians}
The need for flexible, informative, proper priors for mixtures has
been in the statistician's wish list for a long time (e.g. see
\cite{stephens99}).  In this section we derive, from first principles,
the entropic prior for a finite mixture of gaussians. This seems to be
the first informative prior for mixtures, derivable from an objective
principle.  The straight forward application of (\ref{eq:6}) produces
a prior that on the one hand is remarkably close to the conjugate
prior that has been shown most successful in simulations, and on the
other hand, departs from it in a way that has always thought to be
desirable but for which there was no known way to implement.

\subsection{The Model}
We consider a finite mixture of $k$ univariate gaussians with vector of
parameters $\theta = (\mu, \sigma, \omega)$ where $\mu\in \Re^{k}$ is the
vector of $k$ means, $\sigma \in \Re_{+}^{k}$ is the vector of $k$ 
standard deviations and $\omega \in \Delta^{k-1}$ is the mixing probability
vector in the $(k-1)$-dimensional simplex $\Delta^{k-1}$. We use the standard
missing data model for mixtures, i.e., we assume the data is $(x,z)$ has
joint density, for $x\in \Re$ and $z\in \{1,2,\ldots, k\}$ given by,
\begin{equation}
  \label{eq:55}
  f(x,z|\theta) = \omega_{z} N(x;\mu_{z},\sigma_{z})
\end{equation}
where $N(x;a,b)$ denotes the density of the normal distribution with
mean $a$ and standard deviation $b$. The label $z$ is assumed to be
missing from the data so that the marginal density of $x$ has the
desired mixture form,
\begin{equation}
  \label{eq:56}
  f(x|\theta) = \sum_{j=1}^{k} \omega_{j} N(x;\mu_{j},\sigma_{j})
\end{equation}
The trick is to compute the prior on the complete $(x,z)$ likelihood
to disentangle the expression for the entropy.

\subsection{Entropy}
Let $\theta^{o} = (m,s,\omega^{o})$ be the initial guess for $\theta$.
The Kullback number between two distributions (\ref{eq:55}) with parameters
$\theta$ and $\theta^{o}$ is,
\begin{equation}
  \label{eq:57}
  I(\theta:\theta^{o}) = E_{\theta}\left( \log\ 
    \frac{\omega_{z} N(x;\mu_{z},\sigma_{z})}
    {\omega_{z}^{o} N(x;m_{z},s_{z})} \right)
\end{equation}
Computing the expectation by first conditioning on $z$ we obtain,
\begin{eqnarray}
  \label{eq:58}
  I(\theta:\theta^{o}) &=& \sum_{j=1}^{k} \wj \left\{
    I(N(\mu_{j},\sigma_{j}^{2}) : N(m_{j},s_{j}^{2})) + \log\frac{\wj}{\woj}
    \right\} \nonumber \\
    &=& \sum_{j=1}^{k} \wj \left\{
      \log\frac{s_{j}}{\sigma_{j}} + \frac{(\mu_{j}-m_{j})^{2}}{2s_{j}^{2}}
      + \frac{\sigma_{j}^{2}}{2s_{j}^{2}} - \frac{1}{2} 
      + \log\frac{\wj}{\woj} \right\}
\end{eqnarray}
Notice that since $\sum_{j=1}^{k}\wj = 1$ we can take the $1/2$ outside
the sum and it will get absorbed into the proportionality constant for
the entropic prior.

\subsection{Volume Element}
Using (\ref{eq:23}) we can immediately obtain from (\ref{eq:58}) the
entries of the Fisher matrix. The matrix is clearly block diagonal with
gaussian blocks for the $(\mu,\sigma)$ parameters and a multinomial
block for the $\omega$ parameters. From the standard volume elements for
gaussians and multinomials we can write the full volume element as,
\begin{equation}
  \label{eq:59}
  g^{1/2}(\theta)\ d\theta = \frac{d\mu\ d\sigma\ d\omega}
  {\left( \prod_{j=1}^{k}\sigma_{j}^{2}\right)
    \left( \prod_{j=1}^{k} \wj^{1/2} \right)}
\end{equation}
where we are abusing the notation a bit since $d\omega$ must be understood
as $\prod_{j=1}^{k-1}d\wj$ so that $\omega\in \Delta^{k-1}$.

\subsection{Entropic Prior}
Just multiply $e^{-\alpha I(\theta:\theta^{o})}$ with (\ref{eq:59}) to get,
\begin{eqnarray}
  \label{eq:60}
  \pi(\theta|\alpha,\theta^{o}) & \propto &
    \prod_{j=1}^{k} \exp\left\{-\alpha \wj \frac{(\mu_{j}-m_{j})^{2}}
      {2s_{j}^{2}} \right\} \cdot \nonumber \\
    & & \prod_{j=1}^{k} \left( \sigma_{j}^{2} 
     \right)^{\frac{\alpha \wj}{2}-1}\ \exp\left\{
       -\frac{\alpha\wj}{2s_{j}^{2}}\ \sigma_{j}^{2} \right\} \cdot \nonumber\\
    & & \prod_{j=1}^{k}\left(\frac{\woj}{s_{j}}\right)^{\alpha \wj}
    \wj^{-\alpha \wj - 1/2}
\end{eqnarray}
This is a remarkable result. Equation (\ref{eq:60}) says that conditional on
$\omega$ all the components of $\mu$ and $\sigma$ are independent and
independent of each other. Moreover, 
\begin{eqnarray}
  \label{eq:61}
  \mu_{j} |\omega & \leadsto & N\left(m_{j}, \frac{s_{j}^{2}}{\alpha\wj}\right)
   \\
   \label{eq:62}
   \sigma_{j}^{2} |\omega &\leadsto& 
     \mbox{Gamma}\left( \frac{\alpha\wj-1}{2}, \frac{\alpha\wj}{2s_{j}^{2}}
       \right)
\end{eqnarray}
where to obtain (\ref{eq:62}) we have used the change of variables 
$v = \sigma_{j}^{2}$ that produces the jacobian $v^{-1/2}$. The joint 
marginal density of $\omega$ is obtained by integrating (\ref{eq:60})
over $\mu$ and $\sigma$ coordinates obtaining, up to a proportionality
constant that,
\begin{eqnarray}
  \label{eq:63}
  \omega &\leadsto& \prod_{j=1}^{k}\left\{\frac{s_{j}}{\wj^{1/2}}\cdot
    \frac{\Gamma((\alpha\wj-1)/2)}{(\alpha\wj/s_{j}^{2})^{(\alpha\wj-1)/2}}
      \cdot \left(\frac{\woj}{s_{j}} \right)^{\alpha\wj}
        \wj^{-\alpha\wj-1/2} \right\} \nonumber \\
 &\leadsto& \prod_{j=1}^{k} \frac{(\woj)^{\alpha\wj} \Gamma((\alpha\wj-1)/2)}
    {\wj^{(3\alpha\wj+1)/2}}
\end{eqnarray}

\subsection{Posterior}
Let $x^{n}= (x_{1},\ldots, x_{n})$ be the observed data and let
$z^{n}$ be the missing labels. As usual we shake the bayesian wand 
to obtain,
\begin{eqnarray}
  \label{eq:64}
  \pi(\theta,z^{n}|x^{n},\alpha,\theta^{o}) &\propto&
  f(x^{n}|\theta,z^{n}) f(z^{n}|\theta) \pi(\theta|\alpha,\theta^{o})
   \\
  &\propto& \left( \prod_{i=1}^{n}\frac{1}{\sigma_{z_{i}}}
      \exp\left\{\frac{-(\mu_{z_{i}}-x_{i})^{2}}{2\sigma_{z_{i}}^{2}}
        \right\} \right) \left( \prod_{i=1}^{n}\omega_{z_{i}}\right)
       \pi(\theta|\alpha,\theta^{o}) \nonumber
\end{eqnarray}
For $j=1,\ldots,k$ define $k_{j}\in \{1, 2, \ldots, n\}$ by,
\begin{equation}
  \label{eq:65}
  k_{j} = |\{i : z_{i}=j \}|
\end{equation}
and replacing these counts into (\ref{eq:64}) we have,
\begin{equation}
  \label{eq:66}
  \pi(\theta,z^{n}|x^{n},\alpha,\theta^{o}) \propto \prod_{j=1}^{k}
\left\{ \frac{\wj^{k_{j}}}{\sigma_{j}^{k_{j}}} \exp\left\{
\frac{-1}{2\sigma_{j}^{2}}\sum_{i:z_{i}=j}(\mu_{j}-m_{j})^{2}\right\}\right\}
\pi(\theta|\alpha,\theta^{o})
\end{equation}

\subsection{Gibbs Sampler}
Inference is done by sampling $(\theta,z^{n})$ vectors from
the posterior (\ref{eq:66}). To sample from (\ref{eq:66}) we
use Gibbs sampling, i.e. we cycle over the full conditionals for
each of the parameters. Let us use the notation $|\ldots$ to mean
given all the other parameters and the data. Here are the distributions
for each of the terms:

\subsubsection{Conditional for $z^{n}$}
When the vector of mixing probabilities $\omega$ is given the joint
distribution of $z^{n}$ are independent multinomials with $\omega$
as the parameter and independent of everything else. Thus,
for $i=1,2,\ldots, n$
\begin{equation}
  \label{eq:67}
  z_{i}|\ldots \leadsto \mbox{Multi}(\omega_{1},\omega_{2},\ldots, \omega_{k})
\end{equation}

\subsubsection{Conditional for $\mu$}
Here again we have the classic problem of computing the posterior distribution
for the mean of a gaussian given $k_{j}$ independent gaussian observations
when the prior is the conjugate gaussian. Looking at the first term of 
(\ref{eq:60}) and the right hand side of (\ref{eq:66}) we get,
\begin{equation}
  \label{eq:68}
  \mu_{j}|\ldots \leadsto N(a_{j},b_{j}^{2})
\end{equation}
where,
\begin{equation}
  \label{eq:69}
  a_{j} = \frac{\displaystyle \frac{1}{\sigma_{j}^{2}} \sum_{i:z_{i}=j} x_{i}
    + \frac{\alpha\wj}{s_{j}^{2}} m_{j}}
{\displaystyle {\frac{k_{j}}{\sigma_{j}^{2}} + \frac{\alpha\wj}{s_{j}^{2}}}}
\end{equation}
and
\begin{equation}
  \label{eq:70a}
  \frac{1}{b_{j}^{2}} = \frac{k_{j}}{\sigma_{j}^{2}}+ 
    \frac{\alpha\wj}{s_{j}^{2}}
\end{equation}

\subsubsection{Conditional for $\sigma$}
Collecting all the factors with $\sigma_{j}$ from (\ref{eq:66}) and 
the second term from (\ref{eq:60}) we obtain,
\begin{equation}
  \label{eq:70}
  \sigma_{j}|\ldots \leadsto (\sigma_{j}^{2})^{\frac{1}{2}(\alpha\wj-k_{j})-1}
    \exp\left\{ \frac{-1}{2\sigma_{j}^{2}}\sum_{i:z_{i}=j}(\mu_{j}-m_{j})^{2}
      - \frac{\alpha\wj}{2s_{j}^{2}} \sigma_{j}^{2} \right\}
\end{equation}
Now let $v=\sigma_{j}^{2}$, then
\begin{equation}
  \label{eq:71}
  f_{v}(v) = f_{\sigma_{j}}(\sqrt{v}) \frac{1}{2} v^{-1/2}
\end{equation}
Using (\ref{eq:71}) with (\ref{eq:70}) we get,
\begin{equation}
  \label{eq:72}
  v = \sigma_{j}^{2} |\ldots \leadsto
   v^{-a-1} \exp\left\{\frac{-c}{v} - b v\right\}
\end{equation}
where,
\begin{eqnarray}
  \label{eq:73}
  a &=& \frac{1}{2}(k_{j}+1 - \alpha\wj) \\
  \label{eq:74}
  b &=& \frac{\alpha\wj}{s_{j}^{2}} \\
  \label{eq:75}
  c &=& \frac{1}{2}\sum_{i:z_{i}=j}(\mu_{j}-x_{i})^{2}
\end{eqnarray}
We can obtain a useful alternative to (\ref{eq:72}) by doing $u=1/v$ so that
\[ f_{u}(u) = f_{v}(u^{-1}) u^{-2} \]
and we get,
\begin{equation}
  \label{eq:76}
  u = \sigma_{j}^{-2} |\ldots \leadsto
   u^{a-1} \exp\left\{\frac{-b}{u}-c u\right\}
\end{equation}
where $a,b$ and $c$ are given by (\ref{eq:73}), (\ref{eq:74}), 
and (\ref{eq:75}) as before.

The distributions (\ref{eq:72}) and (\ref{eq:76}) are instances of
the so called {\it Generalized Inverse Gaussian} (or GIG for short, see
\cite{devroye86}) distribution. The GIG distribution was first
introduced in relation to hyperbolic distributions in
\cite{barndorff-nielsen77}. It can be shown that,
\begin{equation}
  \label{eq:77}
  \int_{0}^{\infty} u^{a-1}\exp\left\{\frac{-b}{u}-cu\right\}\ du = 
   2 \left(\frac{b}{c}\right)^{\frac{a}{2}} \mbox{BesselK}(a,2\sqrt{bc})
\end{equation}
where the BesselK$(a,x)$ is the modified Bessel function of the third
kind. It is the solution to the differential equation,
\begin{equation}
  \label{eq:78}
  x^{2}y''+xy'-(x^{2}+a^{2})y = 0
\end{equation}
Thus, (\ref{eq:72}) and (\ref{eq:76}) are proper provided that $b > 0$
and $c>0$. When either $b=0$ or $c=0$ (but not both) one of the two
becomes a Gamma.  As it is indicated in \cite{devroye86} the good news
about GIGs is that they are log concave and there are universal
algorithms for generating them. The problem is that the standard off
the shelve algorithm for log concave densities requires the evaluation
of the normalization constant, which in this case is too expensive,
since it involves evaluating BesselK. The following Gamma approximation
provides a solution to this problem.

\subsubsection{Gamma Approximation to GIG}
By computer algebra it is possible to find the parameters of a
Gamma that best fit a given GIG. Let us use the notation, for
$\alpha >0$ and $\beta>0$,
\begin{equation}
  \label{eq:79}
  \Gamma(x;\alpha,\beta) = \frac{\beta^{\alpha}}{\Gamma(\alpha)}
     x^{\alpha-1}\ e^{-\beta x} \ \mbox{for\ } x > 0
\end{equation}
and let, for $a>0, b>0$ and $c>0$,
\begin{equation}
  \label{eq:80}
  G(x;a,b,c) = \frac{1}{Z} x^{a-1}\ \exp\left\{\frac{-b}{x}-c x\right\}
    \ \mbox{for\ } x>0
\end{equation}
where $Z$ is the normalization constant given by the right hand side
of (\ref{eq:77}). We summarize the findings in the next theorem.

\begin{theorem}
\label{th:2}
  The best second order $\Gamma(x;\alpha^{*},\beta^{*})$ approximation to
$G(x;a,b,c)$ is when,
\begin{eqnarray}
  \label{eq:81}
  \alpha^{*} = a \left[1 + \frac{4bc}{\lambda}\right] \\
  \label{eq:82}
  \beta^{*} = c \left[1 + \frac{4bc}{\rho} \right]
\end{eqnarray}
where, 
\begin{eqnarray}
  \label{eq:83}
  \lambda &=& a-1+E \\
  \label{eq:84}
  \rho &=& (a-1)\lambda \\
  \label{eq:85}
  E &=& \sqrt{(a-1)^{2}+4bc}
\end{eqnarray}
\end{theorem}
\proof Here is a summary of what was found with MAPLE.
The function $G(x;a,b,c)$ has a single global maximum at
\begin{equation}
  \label{eq:86}
  x^{*} = \frac{\lambda}{2c}
\end{equation}
Expanding both log likelihoods in Taylor series about $x^{*}$ we get,
\begin{eqnarray}
  \label{eq:87}
  \log\Gamma(x;\alpha,\beta) = A_{0}+A_{1}(x-x^{*})+A_{2}(x-x^{*})^{2}
    + o((x-x^{*})^{2}) \\
  \label{eq:88}
  \log G(x;a,b,c) = B_{0}+0\cdot(x-x^{*})+B_{2}(x-x^{*})^{2}
    + o((x-x^{*})^{2})
\end{eqnarray}
The optimal parameters $\alpha^{*}$ and $\beta^{*}$ are the solution to
the system of equations,
\begin{eqnarray}
  \label{eq:89}
  A_{1}(\alpha,\beta) &=& 0 \\
  \label{eq:90}
  A_{2}(\alpha,\beta) &=& B_{2}(a,b,c)
\end{eqnarray}
\qed \\

The Gamma approximation provided by theorem~\ref{th:2} fits the bulk
of the GIG very well but the tails of the GIG are always heavier.
A few metropolis iterations starting from the gamma approximation
should be used to correct for the light tails.

\subsubsection{Conditional for $\omega$}
Collecting all the factors with $\wj$ from (\ref{eq:66}) and 
all the terms from (\ref{eq:60}) we obtain,
\begin{equation}
  \label{eq:91}
  \omega |\ldots \leadsto \prod_{j=1}^{k} \wj^{\alpha_{j}-1}\ e^{-\beta_{j}\wj}
\end{equation}
where,
\begin{equation}
  \label{eq:92}
  \alpha_{j} = k_{j}-\alpha\wj+1/2 \approx k_{j} + 1/2
\end{equation}
and,
\begin{equation}
  \label{eq:93}
  \beta_{j} = \frac{\alpha}{2}\left[
        \left(\frac{\mu_{j}-m_{j}}{s_{j}}\right)^{2}+ 
        \left(\frac{\sigma_{j}}{s_{j}}\right)^{2} -
        \log\left(\frac{\sigma_{j}}{s_{j}}\right)^{2} + 
        2\log\frac{1}{\woj}\right]
\end{equation}
Notice that $\beta_{j}>0$ and we can use Lemma~\ref{lm:1} again to
find good starting approximations to be corrected with a small number
of metropolis iterations.

\section{Conclusions and Future Work}
We have provided explicit formulas for adding objective prior
information in two general classes of hypothesis spaces: Discrete
probabilistic networks and mixtures of gaussians models.  Many highly
successful models are special cases of BBNs. A partial list lifted
from \cite{murphy2001} include, linkage analysis in genetics, Hidden
Markov Models for speech recognition, Kalman filtering for tracking
missiles, and density estimation for data
compression and coding with turbocodes. It is only natural to expect
improvements in the performance of these methods if there is
available cogent prior information that has not been used. This is
specially true in high dimensional parametric models.

I am currently investigating alternative/complementary methods to MCMC for
performing approximate inference with entropic piors. These include,
the variational bayes approach (see \cite{jordan2001}), and the Expectation
Propagation (EP) method of Minka (see \cite{minka2001}).

\section{Acknowledgments}
This paper was conceived during the summer of 2000 while I was visiting the 
data analysis group at the {\em Center for Interdisciplinary Plasma Science}
(CIPS) \cite{cips}. I would like to thank Volker Dose, Rainner Fischer,
Roland Preuss, Udo von Toussaint and Silvio Gori for many stimulating
conversations.

\bibliography{carlos} \bibliographystyle{maxent95}

\end{document}